\documentclass[12pt,dvips]{article}
\usepackage{amsfonts,amsmath}
\usepackage{graphicx,color}

\newcommand{\la}{\lambda}
\newcommand{\La}{\Lambda}

\newcommand{\cJ}{{\cal J}}

\newcommand{\cN}{{\cal N}}

\newcommand{\tq}{{\tilde q}}

\newcommand{\tLa}{{\tilde \Lambda}}
\newcommand{\tla}{{\tilde \lambda}}
\newcommand{\tm}{{\tilde m}}
\newcommand{\ts}{{\tilde s}}
\newcommand{\tell}{{\tilde \ell}}

\newcommand{\bj}{{\bar j}}

\newcommand{\bL}{{\bar L}}

\newcommand{\ket}[1]{{|#1\rangle}}
\newcommand{\bra}[1]{{\langle#1|}}
\newcommand{\Z}{{\bf Z}}
\newcommand{\R}{{\bf R}}
\newcommand{\C}{{\bf C}}

\newcommand{\p}{{\bf P}}

\newcommand{\n}{\nonumber}
\newcommand{\half}{{1\over 2}}
\newcommand{\del}{\partial}

\newcommand {\eqn}[1]{(\ref{#1})}

\makeatletter
\@addtoreset{equation}{section}

\makeatother

\begin{document}
\vskip 7mm
\begin{titlepage}

\renewcommand{\thefootnote}{\fnsymbol{footnote}}
\font\csc=cmcsc10 scaled\magstep1
{\baselineskip=14pt
\rightline{
\vbox{\hbox{hep-th/0112221}
      \hbox{UT-981}
       }}}

 \vfill
 \baselineskip=20pt
 \begin{center}
 \centerline{\LARGE  Comments on D-branes in Kazama-Suzuki models} 
 \vskip 5mm 
 \centerline{\LARGE and Landau-Ginzburg theories}

 \vskip 2.0 truecm

\noindent{\it \large Masatoshi Nozaki} \\
{\sf nozaki@hep-th.phys.s.u-tokyo.ac.jp}
\bigskip

\vskip .6 truecm
{\baselineskip=15pt
{\it Department of Physics,  Faculty of Science\\
     University of Tokyo\\
     Bunkyo-ku, Hongo 7-3-1, Tokyo 113-0033, Japan}
}
\vskip .4 truecm

\end{center}

 \vfill
 \vskip 0.5 truecm

\begin{abstract}
\baselineskip 6.7mm
We study D-branes in Kazama-Suzuki models by means of the boundary state
description. 
We can identify the boundary states of Kazama-Suzuki models 
with the solitons in $\cN=2$ Landau-Ginzburg theories.
We also propose a geometrical interpretation of the boundary states in
Kazama-Suzuki models.
\end{abstract}

\setcounter{footnote}{0}
\renewcommand{\thefootnote}{\arabic{footnote}}
\end{titlepage}

\newpage

\section{Introduction}
\hspace{6.5mm}
In considering a string compactification, one of the most important
subjects is the physics near a singularity.
Some of cycles vanish near the singularity and various
nonperturbative phenomena take place \cite{strominger,gms}.
To reveal the whole picture of string theory, 
we should further investigate this subject.
In particular, among the many recent studies of string compactification
on singular Calabi-Yau manifolds 
\cite{ov,abks,gvw,gkp,gk,es1,mizoguchi,yamaguchi,nn}, 
Gukov, Vafa and Witten have
argued an interesting proposal \cite{gvw}.
They consider a two dimensional
field theory with four supercharges
constructed in Type IIA string theory on a Calabi-Yau four-fold with an
isolated singularity that 
flows into a non-trivial {\it space-time} superconformal theory at
the infrared fixed point.  
Near an isolated singularity, nonperturbative physics often generates
massless chiral supermultiplets and a superpotential. Then they proposed
that from the non-compact Calabi-Yau four-folds that
are ALE fibrations with appropriate RR-fluxes, one can obtain the
perturbed Kazama-Suzuki models \cite{ks} at level one that have
Landau-Ginzburg descriptions.
In fact, by analyzing the vacuum and soliton structure of each theory,
they have found one to one correspondence between the BPS
D-branes wrapped on a 4-cycle in a non-compact Calabi-Yau four-fold
and the BPS
domain walls in the
Landau-Ginzburg theory that flows to a Kazama-Suzuki model.

Now we would like to discuss what objects in the 
Kazama-Suzuki model correspond to the solitons in the
Landau-Ginzburg theory.
As far as
the closed string physics is concerned, there have been a lot of works to find
the correspondence between the Kazama-Suzuki models  and the
Landau-Ginzburg theories
\cite{lvw,lw,gepner3,intriligator,cv,witten1,flz}.
In particular,
it was shown that the chiral ring of the coset model at level one 
is isomorphic to the de Rham cohomology ring of 
Grassmannian.
It has also been argued that the ring is described in terms of 
the polynomial ring of Landau-Ginzburg theory.

We will consider the correspondence from the viewpoint of
the models that have a boundary on the world-sheet. Then there are
so-called
``boundary states'' in such models that should correspond to
D-branes wrapped on certain cycles.
Now our question is what the boundary states in the
Kazama-Suzuki models describe. 
(Some works have been done for the case of the $\cN=2$ minimal models 
\cite{hiv,lerche,lls,es2,sy} and for the
Kazama-Suzuki models \cite{lwalcher}.)
We will show that the boundary states in the Kazama-Suzuki models
are identified with the solitons in the Landau-Ginzburg theories.
Therefore we propose that the boundary states in the Kazama-Suzuki
models can be thought of as BPS D4-branes 
wrapped on the 4-cycles in a non-compact Calabi-Yau four-fold
with an appropriate RR-flux, or at least they describe a sub-sector of the 
BPS D-branes that sufficiently capture the information of the singularity.
Our result is indeed a first step to understand this relation and
the further investigations are required.

This paper is organized as follows.
We first give a brief review of the 
Kazama-Suzuki model in Section 2 and we
explicitly show the construction of
the boundary states in the Kazama-Suzuki model which indeed
satisfy the Cardy condition \cite{c}.
In Section 3, we summarize the basic facts about the Landau-Ginzburg
theory. In particular, we review how to derive the soliton structure of
the theory which should correspond to the boundary states in
the Kazama-Suzuki model.
Then we find the correspondence between the boundary states in
Kazama-Suzuki model 
and the solitons in Landau-Ginzburg theory in Section 4.
We give two examples there and discuss the geometrical interpretation of 
the boundary states. 
Finally in section 5, we discuss 
some of our results and unsolved problems.
In the appendix, we summarize the characters and 
the S-matrices used in this paper. 

~

\section{Boundary states in Kazama-Suzuki models}
\subsection{Review of the Kazama-Suzuki models}
\hspace{6.5mm}
The Kazama-Suzuki coset models $G/H$ 
are the large class of solvable two dimensional conformal field theories
which possess an $\cN=2$ superconformal symmetry \cite{ks}.
These models were constructed by first 
applying so-called coset ($G/H$) construction to super-Kac-Moody
algebra of group $G$ to obtain $\cN =1$ superconformal algebra. 
Then the conditions under which the $\cN =1$ models actually possess $\cN =2$
superconformal symmetry were examined. 
The result is simple; 
the condition is that the coset manifold $G/H$ is K\"ahlerian.
By way of notation, let $r$ be the rank of $G$, and $g_G$ and $g_H$ are the
dual Coxeter numbers of $G$ and $H$, respectively.
Then the $G/H$ superconformal model can be decomposed as 
\begin{equation}
\frac{\widehat{G}_k\times \widehat{SO}(2d)_1}{\widehat{H}_{k+g_G-g_H}}~,
\end{equation}
where $d=\half{\rm dim}(G/H)$ and 
the $\widehat{SO}(2d)_1$ factor arises from the fermions.
Among the K\"ahlerian models, we consider the special class of the coset
spaces, so-called simply laced, level one, hermitian symmetric spaces 
(SLOHSS). 
In particular, the explicit model we would like to investigate 
here is 
\begin{equation}
\frac{\widehat{SU}(n+k)_1\times \widehat{SO}(2nk)_1}
{\widehat{SU}(n)_{k+1}\times \widehat{SU}(k)_{n+1}\times 
\widehat{U}(1)_{kn(k+n)(k+n+1)}}
\cong
\frac{\widehat{SU}(n+1)_k\times \widehat{SO}(2n)_1}
{\widehat{SU}(n)_{k+1}\times \widehat{U}(1)_{n(n+1)(k+n+1)}}~,
\label{grassmannian}
\end{equation}
where 
the left and right models are equivalent by the level-rank duality 
\cite{ks,lvw,gepner1,ns}.
The central charge of this Kazama-Suzuki model is
$c ={3nk \over k+n+1}$.
From now on, we will consider only the right hand of this model,
namely ``$\C\p^n$ model at level $k$''.

The primary states of this $\cN =2$ coset model are represented as
those of the Kac-Moody algebra of group $G$, $H=H_0\times U(1)$ and
$SO(2d)$.
At first, the highest weights of the Lie algebra associated with $G$
are labeled by
\begin{equation}
\Lambda = \sum_{i=1}^r \ell_i \omega_i~,
\end{equation}
where $\omega_i$ denotes the fundamental weight and
the $\ell_i$'s are all non-negative integers satisfying $\sum
\ell_i \leq k$. 
From now on, we denote the weight as the component form 
$\La =(\ell_1,\ldots,\ell_r)$.
The highest weights of $SO(2d)$ at level 1 are represented by
\begin{equation}
s=0,2,1,-1~,
\end{equation}
which correspond to besic ($s=0$), vector ($s=2$), spinor
($s=1$) and cospinor ($s=-1$) representations, respectively.
Then the basis of the Hilbert space of $\cN =2$ coset model can be
obtained by decomposing ${\cal H}^{\Lambda}\otimes {\cal H}^{s}$ into the
irreducible representations of the group $H$, i.e.
\begin{equation}
{\cal H}^{\Lambda}\otimes {\cal H}^{s}=\sum_{\lambda,m}
{\cal H}^{\Lambda,s}_{\lambda,m}
\otimes {\cal H}^{\lambda,m}~,\label{Hilbert}
\end{equation}
where the $(\lambda,m)$ label the highest weight of $H=H_0\times U(1)$.
Thus the highest weight state is denoted by
$\ket{\Lambda,\lambda,m,s}$.

However not all sub-sectors of the Hilbert space
${\cal H}_{\lambda,m}^{\Lambda,s}$ are independent.
There is a special relation among them.
At first, the characters of the $\cN =2$ coset
model do not vanish if the labels 
satisfy the following selection rule \cite{gepner2,lvw}
\begin{equation}
\frac{|\Lambda|}{n+1}-\frac{|\lambda|}{n}+{m\over n(n+1)}-{s \over 2}=0 
\quad \mbox{mod} \; 1~, \label{selection}
\end{equation}
where $|\Lambda|=\sum_{i=1}^r i \ell_i$ denotes the number of boxes in
the Young tableau corresponding to the weight $\Lambda$ of $SU(n+1)_k$
and $|\lambda|$ denotes the number of boxes corresponding to
the weight $\lambda$ of $SU(n)_{k+1}$ .
There is also another constraint, i.e. the field identification.
The outer-automorphism of the extended Dynkin diagrams 
of $G$ and $H$
will force certain field identifications.
Now let us consider the model (\ref{grassmannian}).
We set the generators of outer-automorphism of 
($\widehat{SU}(n+1)_k,\widehat{SU}(n)_{k+1},
\widehat{U}(1),\widehat{SO}(2d)_1$) 
as $(A,a,h,v)$, where $h=k+n+1$ and $v$ denotes the vector.
Then
the labels $(\Lambda,\lambda,m,s)$ are identified under the action of
$\cJ=(A,a,h,v)$, 
\begin{equation}
(\Lambda,\lambda,m,s)\sim {\cal J}(\Lambda,\lambda,m,s)\equiv
(A(\Lambda),a(\lambda),m+h,s+v)~.
\end{equation}
The order of this generator in the model (\ref{grassmannian}) is $n(n+1)$.
In general, there are some fixed points among the 
${\cal H}_{\lambda,m}^{\Lambda,s}$ space under the field
identifications \cite{schellekens}.
However, for simply laced $G$ at level one there are no fixed point
problems and this is indeed the model we discuss in this paper.

In the case of the model (\ref{grassmannian}),
the conformal dimension $h$ and the $U(1)$
charge $q$ of the primary state $(\Lambda,\lambda,m,s)$ is
\begin{eqnarray}
h&=&{1\over 2(k+n+1)}\left[\Lambda(\Lambda +2\rho_G)
   -\lambda(\lambda +2\rho_H)
   -{m^2\over n(n+1)}\right]
   +g(s) \quad \mbox{mod}\; \Z~,\n\\
q&=&f(s)-{m \over n+k+1}
\quad \mbox{mod}\; 2\Z~,
\end{eqnarray}
where
\begin{eqnarray}
g(s)&=&(0,\half,{d\over 8},{d\over 8}) \quad{\rm for}\quad s=(0,2,1,-1)~,\\
f(s)&=&(0,1,{d\over 2},{d\over 2}-1)\quad{\rm for}\quad s=(0,2,1,-1)
\quad {\rm of} \quad \widehat{SO}(2d)_1~,
\end{eqnarray}
and $\rho_G$ and $\rho_H$ are the
Weyl vectors of $G$ and $H$, respectively.


The chiral primaries ($h=q/2$ mod $\Z$) are labeled by \cite{gepner1}
\begin{eqnarray}
\widehat{SU}(n+1)_k &:& \Lambda =(\ell_1,\ell_2, \cdots, \ell_n)~, \n\\
\widehat{SU}(n)_{k+1} &:& \lambda =(\ell_1,\ell_2, \cdots, \ell_{n-1})~,\\
\widehat{U}(1)_{n(n+1)(n+k+1)} &:& m=\sum_{i=1}^{n}i\ell_i  ~, \n
\end{eqnarray}
and the number of the chiral primaries is  
$
\left(
n+k \atop k
\right)
$.

~


\subsection{Boundary states in Kazama-Suzuki model}
\hspace{6.5mm}
As is well known, 
D-branes can be described either in terms of boundary
conditions of open strings or as boundary states in the closed string channel. 
Boundary conditions for $\cN =2$ superconformal field theory were
investigated in \cite{ooy} and explicit boundary conditions for
Kazama-Suzuki model was given in \cite{stanciu}.
Two types of boundary conditions are possible for $\cN =2$ superconformal
currents 
and are called as the A-type and the B-type conditions.
In the closed string channel, these conditions on the boundary $z=\bar z$ 
of world-sheet are represented as

\noindent{\underline{ A-type boundary condition}:}
\begin{equation}
(T-{\bar T})\ket{B;\eta} = 0 ~, \quad
(G^{\pm} - i \eta {\bar G}^{\mp}) \ket{B;\eta} = 0~,\quad
(J - {\bar J}) \ket{B;\eta} = 0 ~, 
\label{A-type}
\end{equation}
\noindent{\underline{B-type boundary condition}:}
\begin{equation}
(T-{\bar T})\ket{B;\eta} = 0 ~, \quad
(G^{\pm} - i \eta {\bar G}^{\pm}) \ket{B;\eta} = 0~,\quad
(J + {\bar J}) \ket{B;\eta} = 0 ~,
\label{B-type}
\end{equation}
where $\eta =\pm 1$ is the different choice of the spin structure.
From now on, we will consider only A-type boundary condition.

The set of solutions of this condition are known as 
Ishibashi states
\cite{i}
and are 
labeled in the same way as the primary fields, namely
$
\ket{\Lambda, \lambda,m,s}\rangle~,
$
which fulfills the same identification and selection rules\footnote{
D-branes in coset models are recently investigated in a number of 
papers \cite{mms,ga,esark,fs,ishikawa}.
}.
The normalization of the Ishibashi states are defined by the relation 
\begin{equation}
\langle\bra{\Lambda, \lambda,m,s}\tq^{\half(L_0+{\bar L_0}-{c\over 12})}
\ket{\Lambda', \lambda',m',s'}\rangle
=\delta_{\Lambda,\Lambda'}\delta_{\lambda,\lambda'}
\delta_{m,m'}\delta_{s,s'}\chi^{\Lambda,s}_{\lambda,m}(\tq)~,
\end{equation}
where
$\tq=\exp(-2\pi i/\tau)$ is the modulus in the closed string channel
(the modulus of open string channel is $q=\exp(2\pi i\tau)$).
Moreover $\chi^{\Lambda,s}_{\lambda,m}(\tq)$ denotes the character of 
$\cN =2$ coset model and according to (\ref{Hilbert})
the action of modular transformation S on the character is 
\begin{equation}
\chi^{\Lambda,s}_{\lambda,m}(\tq)=\sum_{\Lambda, \lambda,m,s}
S_{\Lambda,\Lambda'}^{SU(n+1)}S_{\lambda,\lambda'}^{SU(n)\dagger}
S_{m,m'}^{U(1)\dagger} S_{s,s'}^{SO(2d)}
\chi^{\Lambda,s}_{\lambda,m}(q)~, 
\end{equation}
where the modular S-matrices of each sector
are summarized in the appendix. 
We should notice here that 
the summation is taken over the primary fields satisfying
the selection rules (\ref{selection}). In other words, it corresponds to
take the summation with respect to all pairs of the labeling with 
the insertion of a Kronecker-delta
\begin{equation}
\delta \equiv \delta^{(n(n+1))}_{n|\La |-(n+1)|\lambda |
+m-{s\over 2}n(n+1),0}~
\end{equation}
of mod $n(n+1)$.
Then from the standard procedure, we can obtain the Cardy states of the
Kazama-Suzuki model (\ref{grassmannian}):
\begin{equation}
\ket{\tLa,\tla,\tm,\ts}=\sum_{\Lambda, \lambda,m,s}\delta
\frac{S_{\tLa,\La}^{SU(n+1)}S_{\tla,\lambda}^{SU(n)\dagger}
S_{\tm,m}^{U(1)\dagger} S_{\ts,s}^{SO(2d)}
}
{\sqrt{S_{0,\La}^{SU(n+1)}S_{0,\lambda}^{SU(n)}
S_{0,m}^{U(1)} S_{0,s}^{SO(2d)}}
}
\ket{\La,\lambda,m,s}\rangle~.
\end{equation}
In fact, we can show that this boundary state in Kazama-Suzuki model 
satisfies the Cardy condition \cite{c}. 
The cylinder amplitude among the boundary states gives the result
\begin{eqnarray}
&&\bra{\tLa', \tla',\tm',\ts'}\tq^{\half (L_0- \bL_0-{c\over 12})}
\ket{\tLa,\tla,\tm,\ts}\n\\
&=&{1\over n(n+1)}
\sum_{\La,\la,m,s}\delta\sum_{\nu=0}^{n(n+1)-1}
\cN_{A^\nu(\La)~ \tLa}^{SU(n+1)}{}^{\tLa'}~
  N_{a^\nu(\la)~ \tla}^{SU(n)}{}^{\tla'}~
\delta^{(n(n+1)(n+k+1))}_{m+(k+n+1)\nu +\tm-\tm',0}~
\delta^{(4)}_{s +2\nu +\ts -\ts',0}~
\chi^{\La,s}_{\la,m}(q)\n\\
&=&
\sum_{\La,\la,m,s}\delta~
\cN_{\La~ \tLa}^{SU(n+1)}{}^{\tLa'}~
  N_{\la~ \tla}^{SU(n)}{}^{\tla'}~
\delta^{(n(n+1)(n+k+1))}_{m+\tm-\tm',0}~
\delta^{(4)}_{s +\ts -\ts',0}~
\chi^{\La,s}_{\la,m}(q)
~,\label{cylinderamp}
\end{eqnarray}
where from the first line to the second line, we used the following fact
that Kronecker-delta could be rewritten as 
\begin{eqnarray}
\delta^{(n(n+1))}_{n|\La|-(n+1)|\la|+m -{s\over 2}n(n+1)}
&=&{1\over n(n+1)}\sum_{\nu =0}^{n(n+1)-1}
e^{{2\pi i \nu\over n(n+1)}(n|\La|-(n+1)|\la|+m -{s\over 2}n(n+1))}\n\\
&=&{1\over n(n+1)}\sum_{\nu =0}^{n(n+1)-1}
e^{2\pi i \nu\left(
{|\La| \over n+1}-{|\la| \over n}\right)}
e^{+{2\pi i m \nu \over n(n+1)}}e^{-\pi i s \nu}~.
\end{eqnarray}
Moreover the action of the outer-automorphism $A$ of $SU(n+1)_k$ 
on the $S$ matrix is (see for example \cite{gepner2})
\begin{equation}
A^{\nu}S_{\tLa, \La}=S_{A^{\nu}(\tLa),\Lambda}=S_{\tLa, \La}
e^{2\pi i \nu {|\La |\over n+1}}~.
\end{equation}
The same equality holds for the $S$ matrix of $SU(n)_{k+1}$.
Finally in the last step of equation (\ref{cylinderamp}), 
we used the character identity 
$\chi^{A(\La),s+2}_{a(\la),m+h}(q)=\chi^{\La,s}_{\la,m}(q)$
and we knew the fact that there was no fixed point in our model.

~


\subsection{$SU(3)/SU(2)\times U(1)$ Kazama-Suzuki model}\label{example}
\hspace{6.5mm}
In this subsection, we explicitly consider the $SU(3)/SU(2)\times U(1)$
Kazama-Suzuki model at level $k$
which corresponds to $n=2$ in (\ref{grassmannian}). 
The result argued here will be used in comparison with the solitons
in the Landau-Ginzburg theory. 

The primary fields in this coset 
model are labeled by the highest weight of $\widehat{SU}(3)_k$,
$\widehat{SU}(2)_{k+1}$, $\widehat{U}(1)_{6(k+3)}$ and $\widehat{SO}(4)_1$.
The set of all highest weights is 
\begin{eqnarray}
\widehat{SU}(3)_k~:&& 
\Lambda=(\ell_1,\ell_2),~~
\;0\leq \ell_1, \ell_2, \;\ell_1+ \ell_2\leq k~,\n\\
\widehat{SU}(2)_{k+1}~:&&  \la =(\la),~~0\leq\la\leq k+1 ~,\n\\
\widehat{U}(1)_{6(k+3)}~: && m,~~ m\in \Z_{6(k+3)} ~,\n\\
\widehat{SO}(4)_1~: && s,~~ s\in \Z_{4}~.
\end{eqnarray}
For the notational convenience, we denote the primary field labeled by
the above highest weights as $((\ell_1,\ell_2),\la, m,s)$. 
As we mentioned before, 
not all the pairs of the labels are independent, but these satisfy
the field identification
\begin{equation}
((\ell_1,\ell_2),\lambda,m,s)\sim
((k-\ell_1-\ell_2,\ell_1),k+1-\lambda,m+k+3,s+2)~,
\end{equation}
and the order of this generator is 6.
This identification is consistent with the selection rule
of the labels
\begin{equation}
{1\over 3}(\ell_1+2\ell_2)-{\lambda \over 2}+{m\over 6}-{s \over 2}=0
\quad {\rm mod}\;1~.\label{selection2}
\end{equation}

The conformal dimension $h$ and $U(1)$ charge of the primary states are
\begin{eqnarray}
h&=&\frac{1}{2(k+3)}\left[
{2\over 3}(\ell_1^2+\ell_1 \ell_2 +\ell_2^2+3\ell_1+3\ell_2)
-{1\over 2}\lambda(\lambda +2)-\frac{m^2}{6}
\right]+g(s)~,\n\\
q&=&f(s)-\frac{m}{k+3}~.
\end{eqnarray}

Then one of the chiral primary state ($h=\half q$) is given by
\begin{equation}
((\ell_1,\ell_2),\lambda,m,s)=((\ell_1,\ell_2),\ell_1,\ell_1+2\ell_2,0)~.
\end{equation}

From now on, let us consider the boundary states in
$SU(3)/ SU(2)\times U(1)$ Kazama-Suzuki model at level $k$.
The Cardy state is 
\begin{equation}
\ket{(\tell_1,\tell_2),\tla,\tm,\ts}=
\sum_{\ell_1,\ell_2,\la,m,s}
\delta~
\frac{S^{SU(3)}_{(\tell_1,\tell_2),(\ell_1,\ell_2)}S^{SU(2)}_{\tla ,\la}}
{\sqrt{S^{SU(3)}_{(0,0),(\ell_1,\ell_2)}S^{SU(2)}_{0, \lambda}}}
{e^{\pi i \tm m\over 3(k+3)}\over \left(6(k+3)\right)^{1\over 4}}
{S^{SO(4)}_{\ts,s}\over \sqrt{S^{SO(4)}_{0,s}}}
\ket{(\ell_1,\ell_2),\lambda,m,s}\rangle~.
\end{equation}
We can immediately read off 
the mass of boundary states 
from the overlap
with the state $\ket{(0,0),0,0,0}\rangle$.
Up to a normalization constant, the mass of the Cardy state is 
\begin{eqnarray}
M((\tell_1,\tell_2),\tla,\ts)&\sim& 
S^{SU(3)}_{(\tell_1,\tell_2),(0,0)}S^{SU(2)}_{\tla ,0}\n\\
&=& \sin{(\tell_1+\tell_2+2)\pi \over k+3}\sin{(\tell_1+1)\pi \over k+3}
\sin{(\tell_2+1)\pi \over k+3}\sin{(\tla+1)\pi \over k+3}~.
\end{eqnarray}
This mass formula will become important, since we use it to identify the
boundary states with the solitons in the Landau-Ginzburg theory.

~


\section{Landau-Ginzburg theory}
\hspace{6.5mm}
In this section, we review some basic facts of the Landau-Ginzburg
theory, particularly of the soliton structure. 
A Landau-Ginzburg theory is believed to flow to a
$\cN =2$ superconformal field theory at a IR fixed point
\cite{martinec,vw}.
The action for the Landau-Ginzburg theory of $n$ chiral superfields 
$\Phi^i$ ($i=1,\ldots,n$) with superpotential $W(\Phi^i)$ is given by 
\begin{equation}
S=\int d^2x\left[
\int d^4\theta K(\Phi^i,{\bar \Phi}^i)
+\half \left( \int d^2\theta W(\Phi^i) + {\rm h.c.} \right)
\right]~,
\end{equation}
where $K(\Phi^i,{\bar \Phi}^i)$ is the K\"ahler potential.

We first explain the basic facts on the 
solitons in the Landau-Ginzburg theory. 
The vacua of this theory are labeled by the critical points of $W$, i.e.
\begin{equation}
\phi^i (x)= \phi^i_*,\quad \del_jW(\phi^i_*)=0 \quad {\rm for \:all} \quad j~.
\end{equation}
The theory is said to be purely massive if all the critical points are 
isolated and non-degenerate. We can then have static domain walls
, or solitons, which interpolate between two different critical points. 
Let us assume that there are non-degenerate critical points;
$\{\phi_a|a=1,\ldots,N\}$.
Then a static soliton $\phi^i(x^1)$ interpolating between 
$\phi^i(-\infty)=\phi_a^i$ and $\phi^i(\infty)=\phi_b^i$ 
has the energy
\cite{fmvw}
\begin{eqnarray}
E_{a,b}&=&\int dx^1\left(
g_{i\bj}{d\phi^i\over dx^1}{d{\bar \phi}^{\bj}\over dx^1}
+{1\over 4}g^{i\bj}\del_iW\del_{\bj}{\bar W}
\right)\n\\
&=&\int dx^1\left|{d\phi^i\over dx^1}-{e^{i\alpha}\over 2}
g^{i\bj}\del_{\bj}{\bar W}\right|^2
+{\rm Re}\left(e^{-i\alpha}(W(\phi_b)-W(\phi_a))
\right)~,
\end{eqnarray}
where $g_{i\bj} = \del_i \del_\bj K$ is the K\"ahler metric and
$e^{i\alpha}$ is an arbitrary constant phase. 
As the energy is independent of the phase $e^{i\alpha}$, if we choose
$e^{i\alpha}=(W(\phi_b)-W(\phi_a))/|W(\phi_b)-W(\phi_a)|$, we can easily
find that there is a lower bound on the energy of the soliton
configuration,
\begin{equation}
E_{ab}\geq |W(\phi_b)-W(\phi_a)|~.
\end{equation}
This is the Bogomol'nyi bound in a supersymmetric theory and
$(W(\phi_b)-W(\phi_a))$ is the central charge in the supersymmetric algebra. 

Now we would like to investigate the soliton structure of Landau-Ginzburg
theory which should have a correspondence with the boundary states in
Kazama-Suzuki model.
In general, the superpotential $W_0(\Phi^i)$ which is related to the
conformal field theory
has isolated quasi-homogeneous singularity at $\Phi^i=0$, i.e.
\begin{equation}
W_0(\lambda^{r_i}\Phi^i)=\lambda W_0(\Phi^i)~,\label{potential}
\end{equation}
and $r_i$ represents the $U(1)$ charge of the lowest component of
$\Phi^i$.
The notion of an isolated singularity means that if we set
$\del_{j}W_0(\Phi^i)=0$ for all $j$, then the solution is at the origin.
The central charge corresponding to (\ref{potential}) is
\begin{equation}
c_W=3\sum_{i=1}^n(1-2r_i)=3(n-2\sum_{i=1}^nr_i)~.
\end{equation}
In fact, the explicit superpotential which corresponds to the model
(\ref{grassmannian}) is known \cite{lvw,gepner3}.
The superpotential of the $SU(n+k)/SU(n)\times SU(k)\times U(1)$ model at
level 1 which is level-rank dual to the $\C\p^n$ model at level $k$
is given by
\cite{lvw}
\begin{equation}
W_0(x_\ell)=x_1^{n+k+1}+\cdots + x_k^{n+k+1}~,
\end{equation}
where $W_0$ can be expressed in terms of $n$ elements 
$\Phi^1, \ldots, \Phi^n$ using the relation that $\Phi^i$ are elementary
symmetric polynomials of $x_\ell$;
\begin{equation}
\Phi^i=\sum_{1\leq\ell_1 < \ell_2<\cdots<\ell_i\leq k}
x_{\ell_1}\ldots x_{\ell_i}~.
\end{equation} 
We can read off the $U(1)$ charge of $\Phi^i$ from this relation and it
is $r_i = {i \over k+n+1}$. Hence the central charge of this
Landau-Ginzburg theory is $c_W = {3nk \over k+n+1}$, which is exactly
the central charge of the corresponding Kazama-Suzuki model.
Moreover the general form of the generating function of
Landau-Ginzburg potential  with respect to the variables $\Phi^i$
is given by Gepner \cite{gepner3}:
\begin{equation}
-\log\left[\sum_{i=0}^{n}(-t)^i\Phi^i\right]
=\sum_{k=-n}^{\infty}t^{n+k+1}W_0^{[n,k]}(\Phi^i)~,
\end{equation}
where we set $\Phi^0=1$.
Then $W_0^{[n,k]}(\Phi^i)$ gives the superpotentilal corresponding to 
the $\C\p^n$ model at level $k$.

However, since the critical points
are multiply degenerate at the origin, we
need to consider relevant perturbations
to discuss the soliton structure.
The perturbations by relevant operators which preserve the $\cN =2$
supersymmetry make them massive field theory.
Although a complete expression of the perturbation is not known, we
perturb the superpotential by the most relevant field,
\begin{equation}
W(\Phi^i)=W_0(\Phi^i)-\mu \Phi^1~,
\end{equation}
where $\mu$ is the perturbation parameter and $\Phi^1$ is the lowest
dimensional chiral superfield.

We can then obtain the soliton structure of the
Landau-Ginzburg theory 
from this superpotential and explicitly draw the 
diagram of the solitons on the $W$-plane.
This diagram is often called the {\it soliton polytopes} of the theory 
\cite{lw}.
The shape of the diagram depends on the models we consider. Therefore
we consider two explicit examples in the next section and 
find a successful correspondence between the diagrams of the
soliton structure and the boundary states in Kazama-Suzuki models.

~


\section{Correspondence between Kazama-Suzuki 
model and Landau-Ginzburg theory}
\hspace{6.5mm}
\label{sub-KSLG}
In this section, we show the correspondence between the boundary states
in the Kazama-Suzuki models and the solitons in the Landau-Ginzburg theories.
We explicitly show two examples here but we believe that this
correspondence is correct for any model.

~


\subsection{Example 1}
\hspace{6.5mm}
Let us consider the simplest example $n=2$, $k=1$ of the model 
(\ref{grassmannian}):
\begin{equation}
\frac{\widehat{SU}(3)_1\times \widehat{SO}(4)_1}
{\widehat{SU}(2)_{2}\times \widehat{U}(1)_{24}}~.\label{level1}
\end{equation}
The primary states in this coset model are listed bellow:
\begin{eqnarray}
((0,0),0,m,s)~,&\quad& m \in \Z_{24}~,\\
((0,0),1,m+3,s)~,&\quad& m \in \Z_{12}~,
\end{eqnarray}
where we only consider NS-sector ($s=0,2$) of $\widehat{SO}(4)_1$. 
There are also other pairs of the labels which represent the primary
fields in the coset model.
However we can show that from the field identification rule, other states
are identified with the fields listed above.
Moreover there is the selection rule (\ref{selection2}) among the
label of each sector and it selects 
the values of $m$ as $m=0~{\rm mod}~6$.
From the result in Section \ref{example}, 
we can read off the mass of the boundary states (up to a proportionality
constant):
\begin{center}
\begin{tabular}{|l|l|c|}
\hline
Primary field & \quad Mass \\
\hline
$\ket{(\tell_1,\tell_2),\tla,\tm,\ts}$ & 
$M((\tell_1.\tell_2),\tla,\tm,\ts)$\\
\hline
$\ket{(0,0),0,\tm,s}$ & \quad $\left(\sin { \pi\over 4}\right)^3
\sin {2 \pi\over 4}$\\
$\ket{(0,0),1,\tm+3,s}$ & \quad $\left(\sin { \pi\over 4}\right)^2
\left(\sin { 2\pi\over 4}\right)^2$\\
\hline
\end{tabular}
\end{center}

Now we return to the solitons in the Landau-Ginzburg theory.
The superpotential that should correspond to the model
(\ref{level1}) is
\begin{equation}
W^{[2,2]}(\Phi^1,\Phi^2)=
{(\Phi^1)^5 \over 5}-(\Phi^1)^3\Phi^2 +\Phi^1 (\Phi^2)^2 -\mu\Phi^1~.
\label{potential1}
\end{equation}
We should comment here that the Landau-Ginzburg theory with 
this superpotential should correspond 
not to $SU(3)_1/SU(2)_2\times U(1)$ model
but to $SU(3)_2/SU(2)_3\times U(1)$ model,
as far as the bulk CFT is considered.
Nevertheless,
the shift of level by one is required to fit the solitons with the
boundary states. 
Such a shift is also observed in the case of
$\cN=2$ minimal model \cite{es2,gkp} which is the
simplest example of Kazama-Suzuki model.
In the minimal model case, they essentially compare the boundary state 
of supersymmetric $SU(2)_N/U(1)$ model with
the soliton of Landau-Ginzburg theory with the superpotential
$\Phi^{N+1}- \mu\Phi$ and obtained the satisfying results
(We will explain this correspondence in more
detail in the next subsection).
We follow this convention and will actually observe 
the successful correspondence
\footnote{
Furthermore this shift is also observed in \cite{lwalcher},
where they discussed that the Landau-Ginzburg potential we should 
use is not the bulk superpotential but the ``boundary superpotential''.
}.

\begin{figure}[tbp]
\begin{center}
\centerline{
\scalebox{1.0}{\includegraphics{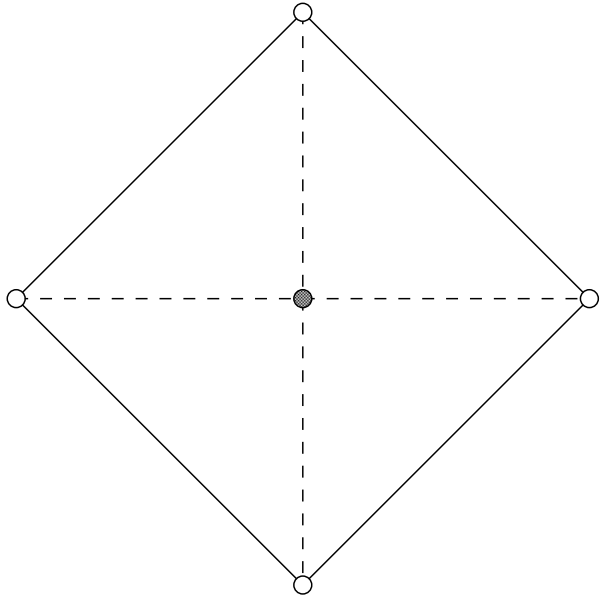}}
}
\caption[]{\small The soliton polytope of Landau-Ginzburg theory with 
the superpotential 
$
W^{[2,2]}(\Phi^1,\Phi^2)=
{(\Phi^1)^5 \over 5}-(\Phi^1)^3\Phi^2 +\Phi^1 (\Phi^2)^2 -\mu\Phi^1
$.
Each circle represents the critical point on the $W$-plane and the line
connecting them is the soliton in the Landau-Ginzburg theory.}
\label{f1}
\end{center}
\end{figure}

Then we can write the polytopes of the Landau-Ginzburg theory
with the superpotential (\ref{potential1}).
There are six critical points on the $W$-plane. 
Among these critical points, four of them form a quadrangle 
up to an overall scaling factor 
(described by white circles in Figure \ref{f1})
and the other two are degenerate at the origin 
(described by filled circles in Figure \ref{f1}).
Then each
straight line which connects any two critical points represents a soliton
in the Landau-Ginzburg theory.
We can find two types of solitons; one is the edges of the quadrangle 
described by solid lines and
the other is the diagonal line of the quadrangle
described by broken lines in Figure \ref{f1}.
The number of solitons is four for edges and two for diagonal lines 
and the ratio of the lengths is $\sin(\pi/ 4)$.
These are exactly the ones which correspond to the boundary states in
the Kazama-Suzuki model listed above.

Now we will give an interpretation of label $m$ 
that will specify the direction
of the soliton in the soliton polytopes.
Let us consider a level-rank duality:
\begin{equation}
\frac{\widehat{SU}(3)_1\times \widehat{SO}(4)_1}
{\widehat{SU}(2)_2\times \widehat{U}(1)_{24}}
\cong
\frac{\widehat{SU}(2)_2\times \widehat{SO}(2)_1}{\widehat{U}(1)_{8}}~.\label{lrduality}
\end{equation}
The right hand side of (\ref{lrduality})
is nothing but the $\cN=2$ minimal model at level 2.
The known fact is that boundary states in this model is related to the
solitons in the Landau-Ginzburg theory \cite{lerche,lls,es2,sy}.
The boundary states in the $\cN=2$ minimal model at level $2$ 
are classified by the
label $(\ell,m',s)$ of the primary fields, where
\begin{eqnarray}
&
0\leq \ell \leq 2~, \quad m'\in \Z_8~,\quad s\in \Z_4~,\quad
\ell +m'+s=0~~{\rm mod}~2 ~,
&\\
&
(\ell,m',s)\sim (2-k,m'+4,s+2)~.
&
\end{eqnarray}
After the field identification, independent boundary states 
in $\cN=2$ minimal model are given as follows:
\begin{eqnarray}
\ket{0,m',s}~, &\quad m'\in 2\Z ~{\rm mod}~8~,&\n\\
\ket{1,m'+1,s}~,&\quad m'\in 2\Z ~{\rm mod}~4~.&
\end{eqnarray}
For simplicity, we set $s=0$,
since it only represents the orientation of the soliton (or brane). 
Then the state $\ket{0,m',0}$ corresponds to the edge line of the
quadrangle and it generates a phase factor $e^{i \pi m' \over 4}$
which gives the direction of the soliton\footnote{
Precisely speaking, we can compute the central charge
from the RR sector of the boundary states.
This central charge is in general complex number and 
the absolute value of the central charge gives the length of soliton.
}. 
Similarly the state $\ket{1,m'+1,0}$ corresponds to the diagonal line of
quadrangle and its phase factor is $e^{i \pi (m'+1) \over 4}$ .

Now let us observe the correspondence between 
the boundary states in $\cN=2$ minimal model and
those in Kazama-Suzuki model.
This can be easily checked by comparing the conformal dimension and the
$U(1)$ charge of each primary state.
The result is 
\begin{eqnarray}
\ket{0,m',0}& \Longleftrightarrow&\ket{(0,0),0,3m',m'}\\
\ket{1,m'+1,0}& \Longleftrightarrow&\ket{(0,0),1,3m'+3,m'+1}
\end{eqnarray}
where $m'$ takes values in $0,2,4,6$ ($3m'=\tm=6\Z$ mod 24).
Thus we were able to 
succeed to identify the boundary states of the Kazama-Suzuki
model and the solitons in Landau-Ginzburg theory. 

On the other hand, there are some solitons that could not be identified 
with the boundary states.
For example, we cannot find a boundary state corresponding to a soliton
that connects the origin with any other critical point.
Moreover when we consider other models at even level $k$,
we always find some critical points at the origin. However we cannot 
identify the boundary states with such solitons which connects the origin
with any other critical point.
On the contrary, as far as we checked, 
the boundary states in Kazama-Suzuki model can
{\it always} be identified with the solitons in Landau-Ginzburg theory
even though we change the level or rank of the
model (\ref{grassmannian}).
This may reflect the fact that we only consider the boundary states
which satisfy the untwisted boundary condition. Therefore other solitons
may be identified with the boundary states satisfying the twisted
boundary conditions \cite{ishikawa} which do not break the 
A-type boundary condition \cite{ooy,stanciu}.
~


\subsection{Example 2}
\hspace{6.5mm}
Next we consider the following coset model at level 2:
\begin{equation}
\frac{\widehat{SU}(3)_2\times \widehat{SO}(4)_1}
{\widehat{SU}(2)_{3}\times \widehat{U}(1)_{30}}~.
\label{example2}
\end{equation}
After using the field identification rule,
inequivalent states which label 
the primary fields in the coset model (\ref{example2}) are given as
follows:
\begin{eqnarray}
&&((0,0),0,m,s)~,\quad
((0,0),1,m+3,s)~,\\
&&((1,1),0,m,s)~,\quad
((1,1),1,m+3,s)~,
\end{eqnarray} 
where we should note that $m=6\Z$ mod 30 from the selection rule.
Then we can calculate the mass of each boundary state.
\begin{center}
\begin{tabular}{|l|l|}
\hline
Boundary state & \quad Mass\\
\hline
$\ket{(\tell_1,\tell_2),\tla,\tm,\ts}$ &
$M((\tell_1,\tell_2),\tla,\tm,\ts)$\\
\hline
$\ket{(0,0),0,\tm,\ts}$ & 
\quad $\left(\sin { \pi\over 5}\right)^3\sin {2\pi\over 5}\;
$\\
$\ket{(0,0),1,\tm+3,\ts}$ & 
\quad $\left(\sin { \pi\over 5}\right)^2
                 \left(\sin {2\pi\over 5}\right)^2
$\\
$\ket{(1,1),0,\tm,\ts}$ & \quad $\left(\sin { \pi\over 5}\right)^2
                 \left(\sin {2\pi\over 5}\right)^2
$\\
$\ket{(1,1),1,\tm+3,\ts}$ & \quad $\sin { \pi\over 5}\,
                 \left(\sin {2\pi\over 5}\right)^3
$\\
\hline
\end{tabular}
\end{center}
Furthermore, by analogy with the previous example
it is natural to propose that
each boundary state in Kazama-Suzuki model
should give a phase factor 
$e^{i\pi m' \over 5}$ or $e^{i\pi (m'+1) \over 5}$,
where $3m'=\tm=6\Z$ mod 30, 
when we compare these boundary states
with the solitons in the Landau-Ginzburg theory.

\begin{figure}[tbp]
\begin{center}
\centerline{
\scalebox{0.9}{\includegraphics{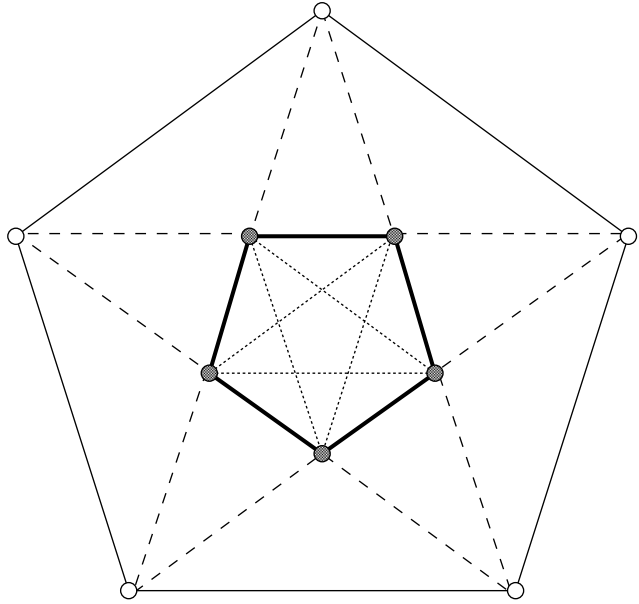}}
}
\caption[]{\small The soliton polytope of Landau-Ginzburg theory with 
the superpotential 
$
W^{[2,3]}(\Phi^1, \Phi^2)={(\Phi^1)^6 \over 6}-(\Phi^1)^4\Phi^2+{2\over 3}
(\Phi^1)^2(\Phi^2)^2-(\Phi^2)^3 -\mu \Phi^1
$.}
\label{f2}
\end{center}
\end{figure}

We then consider the solitons of Landau-Ginzburg theory which should
be identified with the boundary states in the model (\ref{example2}). 
Corresponding Landau-Ginzburg superpotential is
\begin{equation}
W^{[2,3]}(\Phi^1, \Phi^2)={(\Phi^1)^6 \over 6}-(\Phi^1)^4\Phi^2+{2\over 3}
(\Phi^1)^2(\Phi^2)^2-(\Phi^2)^3 -\mu \Phi^1~.
\end{equation}
By using this superpotential, we can calculate the critical points and 
draw a picture of soliton polytope of the theory (see Figure \ref{f2}).

There are ten critical points on the W-plane.
Among them, five points form a small pentagon
(described by filled circles in Figure \ref{f2})
and the other five points form a big pentagon
(described by white circles).
We can find four types of solitons which can be identified as
the boundary states of Kazama-Suzuki model.
The first solitons are represented by the thick solid lines 
which connect two neighboring filled circles and 
they form a small pentagon.
The second solitons are represented by dotted lines 
and they form a small star.
The third solitons are represented by broken lines 
which connect a white circle to a filled circle.
Final solitons are represented by solid lines 
which connect two neighboring white circles and 
they form a big pentagon.
The interesting observation about these four types of solitons is that 
the ratio of the lengths is exactly equal to the mass ratio of the
boundary states. 
Precisely speaking, 
the thick solid lines correspond to
the boundary states $\ket{(0,0),0,\tm,\ts}$,
the dotted lines correspond to
the boundary states $\ket{(0,0),1,\tm+3,\ts}$,
the broken lines correspond to
the boundary states $\ket{(1,1),0,\tm,\ts}$ and finally
the solid lines correspond to
the boundary states $\ket{(1,1),1,\tm+3,\ts}$.
One might be afraid that it is impossible to 
distinct between $\ket{(0,0),1,\tm+3,\ts}$ and $\ket{(1,1),0,\tm,\ts}$
because they give the same mass.
However as we mentioned before, these boundary states should give a phase
factor $e^{i \pi (m'+1)\over 5}$ and $e^{i \pi m'\over 5}$,
respectively and thus the relative angle differs by ${\pi\over 5}$.
By taking account of the phase factors,
we can identify all the boundary states in the model (\ref{example2}) 
with the solitons drawn in the soliton polytope of the Landau-Ginzburg
theory.

On the other hand, there is a problem.
There are ten broken lines
in the soliton polytope as opposed to the fact that there are only
five boundary states $\ket{(1,1),0,\tm,\ts}$ ($\tm=0,6,12,18,24$).
However the disagreement could not be a serious problem since 
among the ten solitons, five of them have independent directions.  
Therefore the boundary states 
may correspond to such independent solitons. 

Finally, we should comment on the $SU(3)/SU(2)\times U(1)$
models at an arbitrary level $k$.
On the Landau-Ginzburg side,
the critical points essentially form 
some regular $(k+3)$-gons on the $W$-plane
and all of the $(k+3)$-gons have the equal center with the
different sizes \cite{flz}.
In particular, when the level $k$ is even, the solitons which have the
smallest mass always correspond to the edges of the smallest $(k+3)$-gon.
(When the level $k$ is odd, we always find the degenerate critical
points at the origin.)
On the other hand,
the boundary states in the Kazama-Suzuki model are labeled by 
$\ket{(\tell_1,\tell_2),\tla,\tm,\ts}$. Now let us consider the state 
$\ket{(0,0),0,\tm,0}$ which have the smallest mass among all the boundary
states.
Then the label $\tm$ takes values in $\tm = 6\Z ~~{\rm mod}~~ 6(k+3)$
and it should give a phase factor 
$e^{\frac{2 \pi i \tm}{6(k+3)}}$.
Thus the states $\ket{(0,0),0,\tm,0}$ can naturally be identified
with the solitons 
represented by the edges of the smallest $(k+3)$-gon on the $W$-plane.
Moreover, we can easily observe that the states 
$\ket{(0,0),\tla,\tm+3\tla,0}$
correspond to the solitons connecting the two arbitrary vertices of 
the smallest $(k+3)$-gon.
Furthermore, as far as we checked,
we can {\it always} find the solitons in the Landau-Ginzburg theory
which correspond to the boundary states
$\ket{(\tell_1,\tell_2),\tla,\tm,\ts}$.

~


\subsection{Geometrical interpretation of boundary states}
\hspace{6.5mm}
Finally, we would like to discuss a geometrical interpretation
of the boundary states in (perturbed) Kazama-Suzuki model.
We will follow the work \cite{gvw,ewy} in this subsection and 
propose that the
boundary states can be identified 
with the D4-branes wrapped on supersymmetric cycles.

For our starting point, we consider type IIA string theory on
$\R^{1,1}\times Y$, where $Y$ is a compact Calabi-Yau four-fold.
The vacuum structure of IIA string compactification is then obtained
by choosing $Y$ and a flux for 
the 4-form field strength $G$ on $Y$.
This flux $G$ originates from the field strength $G=dC$ of
three-form potential $C$ in M-theory on $\R^{1,2}\times Y$ 
and the string theory is derived by circle compactification.
In type IIA string theory, 
this flux is understood as the RR 4-form field.
The difference 
between $\xi\equiv\left[ G\over 2 \pi\right]$ 
and $\xi'$
is classified topologically
by $\xi-\xi'\in H^4(Y;\Z)$.
By Poincar\'e duality, $H_4(Y;\Z)=H^4(Y;\Z)$, there is a 4-cycle
$S\subset X$ which represents an element of $H_4(Y;\Z)$.
The domain walls, or solitons, are then represented by kinks that
interpolate between spatial regions in which $G$-fluxes are different.
Indeed, D4-brane wrapped on a 4-cycle $S \in H_4(Y;\Z)$
appears as a ``particle'' in the world $\R^{1,1}$
and can be regarded as a source of $G$-flux.
If we across the particle, we arrive at a region in which $G$-flux is
different. Thus this D4-brane looks macroscopically like a domain wall,
or a soliton.  
 
For practical reason, it is convenient to omit the part of $Y$ that is far
from the singularity and to consider the non-compact Calabi-Yau
four-fold $X$ that is developing a singularity.
Let us consider Type IIA strings propagating on a singular
Calabi-Yau four-fold 
whose singularity looks like the $A_{N-1}$ singular hypersurface
\begin{equation}
P_N(z_1)+z_2^2+ \ldots + z_5^2 =0~,\label{surface}
\end{equation}
where $P_N(z_1)$ is a polynomial of degree $N$ in $z_1$.
Then for the holomorphic $4$-form $\Omega$ of the non-compact Calabi-Yau
four-fold,
the volume of a supersymmetric 4-cycle $C$ is given by \cite{gvw}
\begin{eqnarray}
V&=&\int_C\left|\Omega\right|\n\\
&=&\left|~\int_IP_N(z_1)~dz_1~\right|~,
\end{eqnarray}
where
the one-dimensional line segment 
$I$ represent the image of the supersymmetric 4-cycle on the $z_1$
plane.
If we define a function $W$ such that
\begin{equation}
dW(z_1) = P_N(z_1)~dz_1~,
\end{equation}
the volume of the cycle $C$ becomes
\begin{equation}
V=\left|~\int_{I_{ab}} dW~ \right| = \left|W(\phi_a)-W(\phi_b)\right|~,
\end{equation}
where $z_1=\phi_a$ and $z_1=\phi_b$ are the end points of segment $I_{ab}$.
Moreover the endpoints of the segment in the 
$z_1$ plane correspond to the critical points in W, namely
$dW=0$ (or $P_N(z_1)=0$). This condition 
is equivalent to find 
critical points in $\cN =2$ Landau-Ginzburg theory with the
superpotential $W$.
Therefore we can identify the D4-branes wrapping on the
supersymmetric cycles 
with the solitons in the Landau-Ginzburg theory 
interpolating between two critical points.
Furthermore from the soliton/boundary state correspondence, 
it seems to be natural to identify the boundary state
with the wrapped 4-branes.

When the polynomial $P_N(z_1)$
represents the deformation of $A_{N-1}$ singularity;
\begin{equation}
P_N(z_1)=(z_1)^{N}-\mu
\quad {\rm or}\quad 
W(z_1)={(z_1)^{N+1}\over N+1}-\mu z_1 ~,
\label{LGpotential}
\end{equation}
the D-branes wrapped on the supersymmetric 
4-cycles of the singular Calabi-Yau four-fold 
are related with
the boundary states in
$\cN =2$ minimal models \cite{es2}.
The system they considered is closely related to the one considered 
in \cite{gvw}.
They considered the holographic dual description of singular Calabi-Yau
compactification of type II string theory in the decoupling limit
by making use of $\cN =2$ minimal models and the 
$\cN =2$ Liouville theory \cite{gkp,es1}.
The soliton structure corresponding to the wrapped D4-branes
are essentially captured from the boundary states in 
$\cN =2$ minimal model.
They evaluated the period of the wrapped branes from 
the boundary states and derived the exact agreement 
with the geometrical calculation including
the scaling behavior.

Now let us consider the Kazama-Suzuki models.
As we mentioned before, the vacuum structure of the theory 
is determined not only by choosing $X$ but also by choosing 
a flux $G$ on $X$.
To specify the theory on the non-compact Calabi-Yau manifold $X$,
we have to fix the flux of $C$-field at the
boundary  $\del X$, i.e. the region near infinity in $X$.
From the quantization condition on the flux \cite{gvw},
the theory is determined by the constant 
flux, $\Phi$, measured at infinity; 
\begin{equation}
\Phi = n + \half \xi^2~,
\label{xi}
\end{equation}
where $n$ is the number of ``string''\footnote{This can be 
thought of as the membrane on $\R^{1,1}\times
S^1$ in M-theory.}. 
However $n\neq 0$ state has massless excitations coming from the
strings, then a model with a mass gap must have $n=0$.
Therefore to get a theory that flows into the massive vacua only
we need to determine minimal $\Phi$. 

As was discussed in \cite{gvw},
when  $X$ is described by the $A_{N-1}$
singular hypersurface \eqn{surface},
then the homology $H_4(X;\Z)$ is naturally
identified with the root lattice $\Gamma$ of $SU(N)$ and the
intersection form is the Cartan matrix of the group.
While the full set of non-trivial $G$-flux $\xi$ are classified by 
$H^4(X;\Z)$ which  can be identified with the weight lattice $\Gamma^*$
of $SU(N)$.
Moreover the possible values of $\xi$ 
modulo the changes due to crossing the domain walls are
classified by 
$\Gamma^* / \Gamma$, that is isomorphic to
the center $\Z_N$ of $SU(N)$ \cite{gvw}.
The distinct elements of $\Gamma^*/\Gamma$ are defined by the {\it
congruence class} (see for example, Chapter 13 of \cite{cft}).
When the center is equal to $m$ ($m=0,1,\ldots,N-1$), 
$\xi$ must be the weight of
$m$-fold anti-symmetric tensor product of fundamental representations
of $SU(N)$ to minimize $\half\xi^2$ in \eqn{xi}. 
If we denote the representation by $R_m$, then the number of choice of 
$\xi$ is then the dimension of $R_m$, i.e. 
${\rm dim}(R_m)=\frac{N!}{m!(N-m)!}$.

Now let us consider a few examples. At first, when $m=1$, then
the dimension or the number of vacua becomes ${\rm dim}(R_1)=N$,
which is exactly the number of vacua of Landau-Ginzburg theory with the
superpotential \eqn{LGpotential}.
Secondly, when $m=2$, then the dimension becomes 
${\rm dim}(R_2)=\half N(N-1)$.
In particular, when $N=k+3=4,5$, the number of vacua is $6,10$, which are
those of vacua in Landau-Ginzburg theory discussed in Example 1 and 2.
Therefore we propose that 
the solitons which represent D4-branes
wrapped on the supersymmetric 4-cycles can naturally be identified 
with the boundary states in Kazama-Suzuki model.
In addition, the boundary states we considered satisfy the A-type boundary
condition \cite{ooy,stanciu} and they essentially correspond to the
D-branes wrapping on the middle dimensional cycles, i.e. supersymmetric
4-cycles. This fact also supports our interpretation.

However, the crucial difficulty is that
the Landau-Ginzburg theory describes the {\it space-time} CFT, on the other
hand our Kazama-Suzuki model should describe the {\it world-sheet}
CFT to give a correct meaning of boundary state. 
This difference might be related with the shift of level by one
when we compared the solitons with the boundary states.
Instead of dealing such a space-time CFT directly,
the holographic dual description of 
superstring theory on a non-compact Calabi-Yau
four-fold is proposed in \cite{gkp,es1} by the $\cN =2$ minimal model 
and the $\cN =2$ Liouville theory.
Then the deformation parameter $\mu$ of the
Landau-Ginzburg theory \eqn{LGpotential} is identified with the
parameter of perturbation by the cosmological constant operator in
the $\cN =2$ Liouville theory.
Likewise, the Kazama-Suzuki model may appear in the 
holographic dual description of such a Calabi-Yau compactification 
with the RR-flux.
We do not know whether this dual description is valid for any value of
the RR-flux. However the soliton structure sufficiently
characterize the theory, and hence the strong relation with the boundary
states in Kazama-Suzuki model would give the natural interpretation that
the boundary states describe the BPS D4-branes.

We do not know, so far, 
any non-compact Calabi-Yau compactification by use of 
Kazama-Suzuki models.
It would be interesting to construct the consistent modular invariant theory
combined in the $\cN =2$ Liouville theory or its generalization. 

~


\section{Summary and discussions}
\hspace{6.5mm}
We studied the correspondence between the
boundary states in Kazama-Suzuki models and the solitons
in Landau-Ginzburg theories.
We have found the successful correspondence by comparing  
the mass of boundary state with the one of soliton including the
direction on the $W$-plane.
We further discussed a geometrical interpretation of the 
boundary states in Kazama-Suzuki
models and proposed that the boundary states are naturally
identified with the D4-branes wrapping on a supersymmetric cycles
in a non-compact Calabi-Yau four-fold with an appropriate RR-flux.

On the other hand, there is an open problem that
the number of the solitons in the Landau-Ginzburg theory
is much larger than that of the
boundary states in Kazama-Suzuki models. 
This problem occur even
if we change the level or rank of the models.
However we only consider the boundary states which satisfy the untwisted
boundary condition. Hence it is interesting to consider the boundary
states which satisfy the twisted boundary conditions
\cite{ishikawa,stanciu}.
 
Moreover it is important to clarify the relation between the
space-time CFT which is realized as the IR fixed point of two
dimensional field theory obtained from a singular
Calabi-Yau four-fold compactification
and the world-sheet CFT described by the Kazama-Suzuki model.
In this paper, we take a position that the the solitons in the
space-time Landau-Ginzburg theory are related to the boundary states
in the world-sheet Kazama-Suzuki model in a sense of holography
\cite{gkp,es2}.
To completely understand this relation, 
it is important to construct the modular invariant partition
function which satisfy the condition of critical dimension
and evaluate the periods of the supersymmetric cycles.
This construction is not well understood so far and may be achieved by
combining with the Liouville theory or its generalization \cite{blnw}. 

Finally in this paper, we have only discussed the Grassmannian models,
in particular we explicitly checked the correspondence between
the boundary states and the solitons in 
a few simple models. 
It is therefore 
important to check this correspondence for other levels or ranks
to understand the general correspondence.
It would be also interesting to check
this correspondence for other SLOHSS models \cite{ks,lw}.
Moreover, as in the $\cN =2$ minimal model cases \cite{hiv,mms},
it is valuable to discuss the geometrical meaning of 
intersection forms in Kazama-Suzuki models \cite{lwalcher}.

~ 

\section*{Acknowledgments}
\hspace{6.5mm}
I am very grateful to my advisor, Kazuo Fujikawa for teaching me and
encouraging me. 
I am also thankful to thank Hiroyuki Fuji, Masaaki Fujii,
Yasuaki Hikida, Michihiro Naka, Kazuhiro Sakai and Yuji Sugawara
for useful discussions and comments.

\appendix

\section{S-matrices in Kazama-Suzuki model}
\hspace{6.5mm}
In the appendix, we summarize S-matrices appeared in the Kazama-Suzuki
model.

~

\noindent{{\underline{{\it $SU(n+1)$ WZW model}}}

~

Let us denote the highest weight of $SU(n+1)$ at level $k$
as $\La$.
Then modular transformation S-matrix for $SU(n+1)$ WZW model at level $k$
can formally be written in the following way:
\begin{equation}
S_{\La,\tLa}^{SU(n+1)}=\frac{i^{\half n(n+1)}}{\sqrt{n+1}(k+n+1)^{n/2}}
\sum_{w\in W}\epsilon(w)e^{-2\pi i w(\La+\rho)(\tLa+\rho)/(k+n+1)}~,
\end{equation}
where $W$ denotes the Weyl group of $SU(n+1)$ and $\rho$ is the Weyl
vector.

~

\noindent{{\underline{{\it $SU(3)$ WZW model}}}

~

We denote the highest weights of $SU(3)$ at level $k$  as $\la$
and it is labeled by 
\begin{equation}
{\cal P}_+^k = \{\Lambda=\ell_1 \omega_1+\ell_2 \omega_2|\ell_1, \ell_2 \in \Z
, \;0\leq \ell_1, \ell_2, \;\ell_1+ \ell_2\leq k
\}~,
\end{equation}
where $\omega_1$ and $\omega_2$ are fundamental weights of $SU(3)$.
Then a explicit form of the
modular S-matrix for $SU(3)$ WZW model at level $k$
can be written in the following way:
\begin{eqnarray}
(S^{SU(3)})_{\lambda, \lambda '}\!\!&=&\frac{-i}{\sqrt{3}(k+1)}
\left\{
e^{{-2\pi i\over 3(k+3)}[2(\ell_1+1)(\ell_1'+1)+(\ell_1+1)(\ell_2'+1)
+(\ell_2+1)(\ell_1'+1)+2(\ell_2+1)(\ell_2'+1)]}\right.\n\\
\!\!&&
\quad\quad\quad\quad\;\;\,
+e^{{-2\pi i\over 3(k+3)}[-(\ell_1+1)(\ell_1'+1)-2(\ell_1+1)(\ell_2'+1)
+(\ell_2+1)(\ell_1'+1)-(\ell_2+1)(\ell_2'+1)]}\n\\
\!\!&&
\quad\quad\quad\quad\;\;\,
+e^{{-2\pi i\over 3(k+3)}[-(\ell_1+1)(\ell_1'+1)+(\ell_1+1)(\ell_2'+1)
-2(\ell_2+1)(\ell_1'+1)-(\ell_2+1)(\ell_2'+1)]}\n\\
\!\!&&
\quad\quad\quad\quad\;\;\,
-e^{{-2\pi i\over 3(k+3)}[-(\ell_1+1)(\ell_1'+1)-2(\ell_1+1)(\ell_2'+1)
-2(\ell_2+1)(\ell_1'+1)-(\ell_2+1)(\ell_2'+1)]}\n\\
\!\!&&
\quad\quad\quad\quad\;\;\,
-e^{{-2\pi i\over 3(k+3)}[2(\ell_1+1)(\ell_1'+1)+(\ell_1+1)(\ell_2'+1)
+(\ell_2+1)(\ell_1'+1)-(\ell_2+1)(\ell_2'+1)]}\n\\
\!\!&&
\quad\quad\quad\quad\;\;\,
\left.
-e^{{-2\pi i\over 3(k+3)}[-(\ell_1+1)(\ell_1'+1)+(\ell_1+1)(\ell_2'+1)
+(\ell_2+1)(\ell_1'+1)+2(\ell_2+1)(\ell_2'+1)]}
\right\}
\end{eqnarray}

\noindent{\underline{{\it $\widehat{U}(1)_L$ model}}}

The modular S-matrix for $U(1)$ model at level $L$
is given by
\begin{equation}
S_{m,\tm}^{U(1)}={1\over \sqrt{L}} e^{-{2 \pi im\tm\over L}}
\end{equation}
where $m,\tm \in \Z_{L}$.

~

\noindent{\underline{{\it $\widehat{SO}(2n)_1$ model}}}

~

In general, 
the character of $\widehat{SO}(2n)_1$ is given by
\begin{eqnarray}
\chi_0^{SO(2n)}=\half
\left[
\left(\theta_3\over \eta \right)^{n} + \left(\theta_4\over \eta \right)^{n}
\right]~, &\quad&
\chi_2^{SO(2n)} =\half
\left[
\left(\theta_3\over \eta \right)^{n} - \left(\theta_4\over \eta \right)^{n}
\right]~,\n\\
\chi_1^{SO(2n)}=\half
\left[
\left(\theta_2\over \eta \right)^{n} +\left(-i\theta_1\over \eta \right)^{n}
\right]~, &\quad&
\chi_{-1}^{SO(2n)}=\half
\left[
\left(\theta_2\over \eta \right)^{n} -\left(-i\theta_1\over \eta \right)^{n}
\right]~,\n
\end{eqnarray}
or
\begin{eqnarray}
\chi_0^{SO(2n)} +\chi_2^{SO(2n)} =\left(\theta_3\over \eta
\right)^{n}~, &\quad&
\chi_0^{SO(2n)} -\chi_2^{SO(2n)} =\left(\theta_4\over \eta 
\right)^{n}~,\n\\
\chi_1^{SO(2n)} +\chi_{-1}^{SO(2n)}=\left(\theta_2\over \eta 
\right)^{n}~, &\quad&
\chi_1^{SO(2n)} -\chi_{-1}^{SO(2n)} =\left(-i\theta_1\over \eta 
\right)^{n}~.\n
\end{eqnarray}
The modular 
S-matrix of character $\chi_s^{SO(2n)}(\tau)$ ($s=0,2,1,-1$) is 
\begin{equation}
S_{s,\ts}^{SO(2n)}=
\half \left(
\begin{array}{cccc}
1 &  1 &  1      &  1       \\
1 &  1 & -1      & -1       \\
1 & -1 &  i^{-n} & -i^{-n}  \\
1 & -1 & -i^{-n} &  i^{-n}  
\end{array}
\right)~.
\label{d=4}
\end{equation}
For $n={\rm odd}$ case, the above matrix can be simply expressed as 
\begin{equation}
S_{s,\ts}^{SO(2n)}=\half e^{-{n\pi i \over 2}s \ts}~.
\end{equation} 

~

\end{document}